# Modelling of a captive unmanned aerial system teledetecting oil pollution on sea surface

F. Muttin
*Engineering School EIGSI La Rochelle, France.*

**Abstract**
Recent major oil-spills were tracked using observations with sufficient altitudes over the sea surface, to detect oil slick locations. For oil-spill responders, we propose a captive Unmanned Aerial System, UAS acting like a periscope over a ship or supply vessel. The system is composed of an umbilical deployed from ship deck, and there are few studies that have examined elasticity within cable dynamic during take-off or landing (TOL) and normal flight phases. Therefore, the safest approach for the control-commands of the system is through umbilical dynamic modelling. We give a time-dependant finite-element formulation, using improved elastic non-linear cable elements. Two kinds of boundary condition, natural or essential, are discussed for roll-in or roll-out of the umbilical. A numerical convergence and a validation with an exact solution are provided, using two examples for the flight parameters. Finally, sensitivity of the model potentially extends its capacity for the system equilibrium prediction, under wind primary influence.

*Keywords: Unmanned aerial system, umbilical, oil-spill, detection, dynamic modelling, winch, maritime pollution.*

## 1 Introduction

During 2010, and the Deep-Water Horizon crisis in the Gulf of Mexico, numerous aerial means were used to detect oil pollution locations on the sea surface [1]. Already, in 2009, from the northwest Australian coast, and the Timor Sea, aerial sea surveillance and oil trajectory modeling were used, during the Montara wellhead platform spill [2].

An observation point altitude corresponding to ship decks is too low to detect oil slick far away from the ship. With altitude 5 m, the detection distance is 300 m. With altitude 150 m, the distance is expected to reach 1500 m.

The aerial system described here is a small drone containing a camera. It is connected at ship deck with a light stiff cable, named umbilical. The system flies using self-stator, and as rotor an electrical brushless engine powering blades. The umbilical contains an electric conductor to carry energy between the ship and the engine. Such unmanned aerial vehicle, UAV also usually named system, UAS [3] acts like a periscope.

Our objective focuses on a dynamic modeling of umbilical. For that, we use a dynamic cable finite-element. It takes into account umbilical elastic strain. The stability, the altitude control and command of the system can be studied more precisely together a dynamic modeling [4]. To simplify the problem, we will consider regular ship motion on sea surface, uniform atmospheric flow over sea, and we assume constant lift-thrust force from the drone.

Satellite images, for example using CleanSeaNet [5] during routine detection, are not treated here. Another drone application concerning tactical UAS [6], for oil terminal protection as example, is not considered here [7].

The paper is organized as follows. First, the system definitions are given. Secondly, we define the principle of the dynamic modeling. After, we describe the finite-element approximations of the problem considering take-off or landing. Finally, we give several numerical results and theirs interpretations for normal flight.

## 2 Materials

The system specifications are examined in this section. It concerns ship, winch, umbilical, camera, and drone as well as marine and coastal environments. The ship using the system can be a fish boat, an oil recovery vessel, a boat supply or an assistance ship of the Navy.

The system can be used for oil location research, or ship guiding for oil recovery [8]. Following oil detection route, the ship velocity can reach 15 knots. During oil recovery, using trawling net, sweeping arm or towed boom as examples, the ship velocity must be less than or equal to 0.7 knot [9].

Many maritime workers were mobilized on ship deck during oil recovery. The safety of workers must be guarantee during the take-off or landing of the system. For that reason, a launching platform is adjacent to the system, which connects it to the ship deck.

The winch is only a roller rotating on a horizontal axis parallel to the ship deck surface. It allows appropriate take-off and landing directions from the launching platform. The roller diameter is around a half-meter. It permits to be large enough to guarantee umbilical conservation and accurate order of umbilical stacking during roll-in and roll-out. The roller width must be large enough, around a half-meter, reducing the number of umbilical roll-in layers, and decreasing the curvilinear velocity variation of the umbilical during unwinding or rewinding.

The umbilical is composed of four subsets; an electric conductor, an electric insulation, a high modulus fibber, and an envelop protection. The high voltage electric current is around 450 V. It permits to reduce significantly the energy loss resulting from linear electric resistance (Joule effect). The mechanical fibber is synthetic and it is made of ultra high molecular weight polyethylene, UHMWPE having a specific gravity of 0.95. The total curvilinear mass can reach 1 kg for 100 m.

The cameras generally used for oil detection and observation on the sea surface were constructed for visible ray, infrared red or laser; and some of them are named light detection and ranging, LIDAR [10]. For mass optimization, light camera for the visible spectrum, approximately in 400 to 700 nm [11-12] is possible. The camera is attached to a double axes rotating pod. The vertical axis runs 360° around the ship. The second axis corresponds to the azimuth, and runs in the interval [0°, 90°]. Between 88° and 89°, it permits oil detection far away from the ship. The control-command of the camera, the signal transmission and treatment, sensors and pilots, are not described here.

The drone uses a vertical take-off and landing, VTOL architecture [13]. Only the lift-thrust force sustains umbilical and drone-camera weights. The drone altitude control-command uses actions on engine regime and on two orthogonal flaps straighten the twisted exhaust flow. The drone mass is around 5 kg, without umbilical. The annular structure of such VTOL is made of composite sandwich material, permitting mass-stiffness improvement. The propeller choice focuses on two or three blades alternative. It will be not detailed here.

The environment of the system is subjected to sea state, atmospheric flow and pollution nature. The geometry taken by oil layer during windy condition is fragmentation of the pollution into parallel lines. As a consequence, ship route must be downwind or headwind during oil recovery. The interaction of ship response from sea state and manoeuvring is not detailed.

The atmosphere-ocean interaction, the ship deck motion and environment, including spill response materials, ship antennas, render the take-off and the landing difficult for the controls-commands of winch, flaps and engine. The consequences of sea states, wind turbulences and wind gradients are not described here.

## 3 Method

The dynamic modelling is based on the fundamental mechanical principle applied to the drone and the umbilical, more the linear elasticity of the cable undergoing large displacements [13]. The boundary condition at bottom end-point can be a constant force tangent to the umbilical curve, or a prescribed displacement. The head-point boundary condition is the drone lift-thrust force tangent to the curve.

A first external force is determined by the aerodynamic pressure $F_v$ of the air overlapping the umbilical and the drone. It is defined in term of the velocity discontinuity square between umbilical and wind. The second external force is the body force $P$ of the umbilical and the drone.

A first internal force is the lift-thrust force ($F_T$, $F_L$) provided by the blades through the straighten twisted air exhaust. A second internal force is the umbilical elastic strain tension $T$. The strain is defined between initial and dynamic states by the Euclidian norm of the curvilinear unit tangent to the umbilical. The last internal force is the winch action $R$ on the bottom end-point.

The aerodynamic drag coefficient of the umbilical section is assumed constant and independent of umbilical curvature. This coefficient takes into account a simple projection rule, based on the angle between horizontal wind flow and umbilical normal vector.

Figure 1 shows two finite-elements, external and internal forces. The assemblage of elementary forces in nodal vectors is equally indicated.

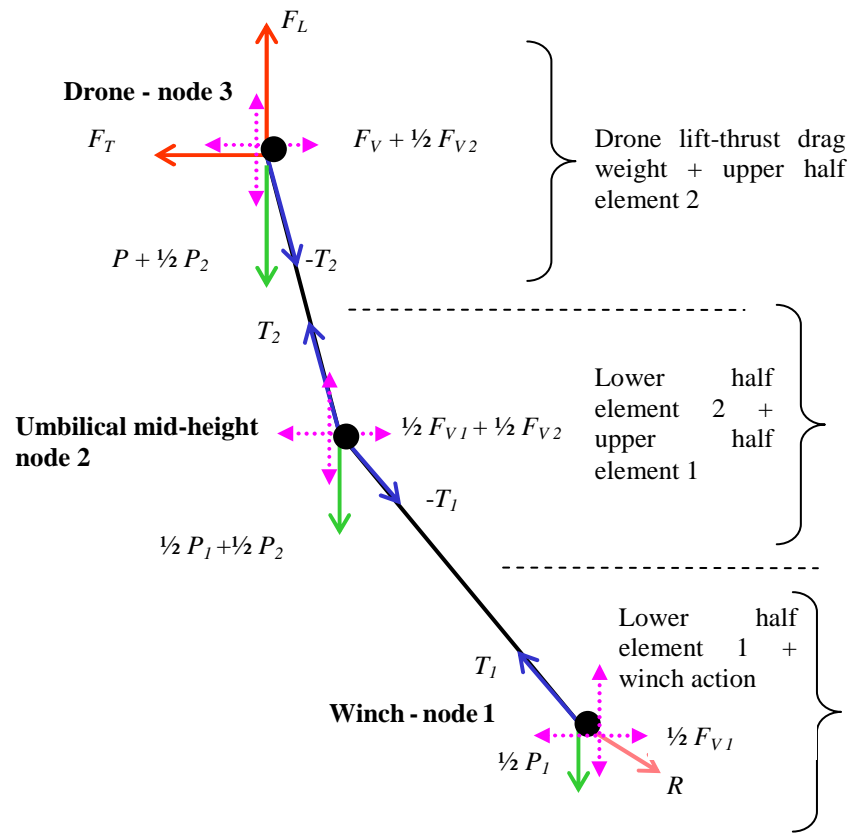

Figure 1: External and internal forces of the UAV system.

The umbilical material is supposed linear and elastic. The Young modulus is supposed constant during time.

## 4 Approximation of the problem

A finite-element mesh permits a geometric approximation of the problem. Additionally, a time integration scheme delivers an approximation of the transient solution. When the umbilical is fully deployed, during normal flight phase, the finite-element mesh formulation is Lagrangian. Each node and each element correspond to a physical part of the umbilical. When the umbilical is not fully deployed, during take-off or landing phases, two geometrical formulations are given for umbilical approximation.

A first formulation uses a decomposition of the finite-element mesh in two subsets. The first one corresponds to a flying deployed part, and the second is based on a sliding part, assumed to move vertically in a virtual well. The boundary between these subsets moves during time along the umbilical curve. It corresponds also to the winch position on ship deck; taken as the virtual well head. Each finite-element located at the bottom is subjected to the solely internal strain action. The internal tension depends on elastic strain. At the bottom end-point of this subset, we define a constant force as the winch action. This formulation is illustrated in figure 2.

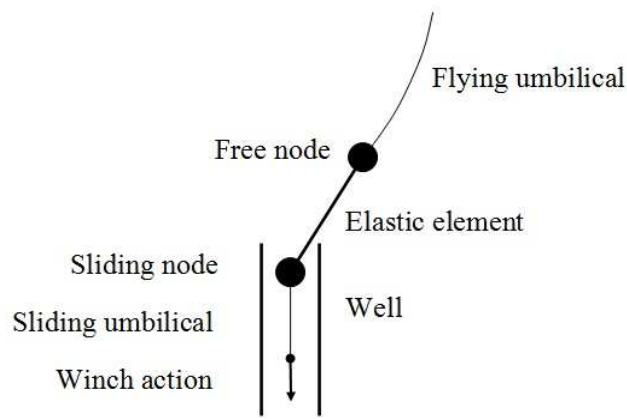

Figure 2: The Lagrangian formulation.

The second formulation is Eulerian-Lagrangian. To simulate the roll-in and roll-out of the umbilical, several nodes are disposed at the winch position. The others nodes correspond to an umbilical flying part. The nodes located near the winch are fixed, while the others are free. Elastic cable elements are defined upon the flying part. Between the two nodes located near the umbilical bottom, we define a constant force element. Its tension corresponds to the winch action. Its length depends on the umbilical tension at its upper boundary. The bottom node of this element is one of the fixed nodes of the winch. When the equivalent unstrained length of this element reaches a threshold value, corresponding to an unstressed equivalent length of the umbilical, the element nature does change into classical elastic element. Therefore, another constant force element is defined between its bottom node and a supplementary winch node. The definitions of the Eulerian-Lagrangian formulation are presented in figure 3.

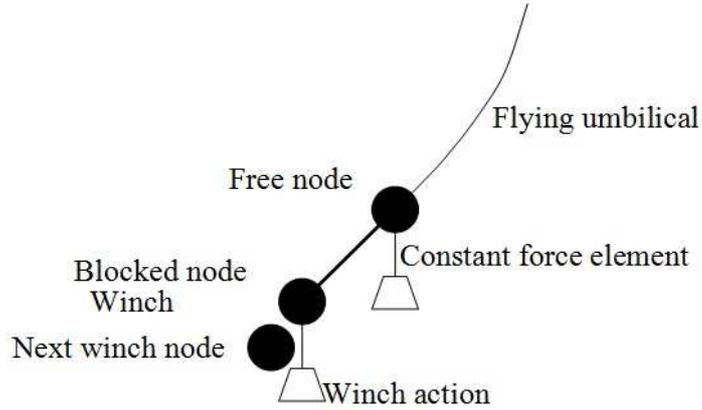

Figure 3: The Eulerian-Lagrangian formulation.

Eulerian-Lagrangian formulation must avoid zero-length for constant force element, by assuming an infinitesimal distance between winch nodes. During a take-off phase, with the Lagrangian formulation, the nodes become successively free, when they come out of the virtual well. During time, when the umbilical is not perfectly vertical, this succession introduces an approximation in the continuous problem of deployment.

Practically, the number of finite-elements is under or equal 10. It suggests reduction of computational time. A maximum of 11 nodes is assumed for such modelling.

## 5 Results

The computations are 2D ($x, z$), where $x$ is the horizontal (oriented in the ship direction) and $z$ is the vertical.

### 5.1 Finite-element convergence

First, we present the results obtained with decreasing finite-element sizes. It permits to estimate the numerical scheme convergence. Here, the finite-element mesh uses $n = 2$, 6 and 8 finite-elements. The parameters of the physical problem are: umbilical total length $L_0 = 150$ m, umbilical diameter 8 mm, (module Young) x (section) of the structural fibber 24 kN, sub-systems masses: drone 3 kg, umbilical 8.7 kg (58 g/m), relative wind velocity 4 m/s (horizontal), constant aerodynamic force from the drone source: vertical lift $F_L$ 238 N, horizontal thrust and drag $F_T+F_V$ 58 N. The initial condition is fully deployed vertical umbilical in zero-tension above the winch. The simulation duration is 30 s. It is assumed sufficient to reach steady state. The ship is motionless along the vertical. For that case, table 1 permits to validate numerical convergences of both geometry and tension.

Table 1: Convergence of the finite-element approximations.

| $n$ | 2 | 6 | 8 |
|---|---|---|---|
| Drone (m) | | | |
| $x$ | -43.09 | -43.61 | -43.56 |
| $z$ | 144.76 | 144.55 | 144.57 |
| Umbilical mid-height (m) | | | |
| $x$ | -23.3 | -24.0 | -24.0 |
| $z$ | 71.8 | 71.5 | 71.5 |
| Viscous force resultant on the umbilical (N) | | | |
| $F_x$ | 3.65 | 3.64 | 3.60 |
| $F_z$ | -0.006 | -0.006 | -0.006 |
| Umbilical tension (N) | | | |
| Head | 192 | 205 | 207.6 |
| Bottom | 148 | 135 | 133 |
| Umbilical elongation (m) | | | |
| $\Delta L_0$ | 0.080 | 0.081 | 0.085 |

The result differences are small. With $n = 6$ and 8 finite-elements, the approximation seams the most acceptable for both geometry and tension.

### 5.2 Analytic validation

The following case is based on a take-off along the vertical. For this purpose we assume zero wind, zero horizontal drag and vertical thrust-lift. The forces being vertical, the drone trajectory depends only on its altitude $z$.

Such assumptions give a simple model for the continuous roll-out of the umbilical. Considering umbilical maximal length $L$, and a take-off along the winch vertical, the flying system mass increases as the time-dependant length of the umbilical. The umbilical roll-out length in [0, $L$], yields for the system mass

$$m = m_0 + \underline{m_{lin}} \cdot z \tag{1}$$

where $m_0$ is the constant drone mass, $m_{lin}$ is the curvilinear umbilical mass in deformed state. The trajectory being on the winch vertical, the umbilical strained length equals the drone altitude and the umbilical mass reaches $\underline{m_{lin}} \cdot z$ .

The lift-thrust force $F_L$, the vertical drag in $v^2$, where $v$ is the drone velocity $|\dot{z}|$, and the gravity acceleration $g$, give the dynamic equation

$$\ddot{z} = \frac{F_L}{m_0 + m_{lin}z} - \frac{k_z \dot{z} |\dot{z}|}{m_0 + m_{lin}z} - g \qquad (2)$$

where $k_z$ is the vertical drag coefficient of the drone. The exact solution of eqn (2) can be estimated numerically to favour the comparison with our numerical scheme results.

The parameters of the physical problem are: $m_0 = 4$ kg, $m_{lin} = 15.10^{-3}$ kg/m, $L = 120$ m, $k_z = 0.3$ Ns²/m, $F_L = 100$ N. In figure 4, the drone altitude during umbilical roll-out with the above continuous model, is compared with the result of the discrete model using 10 finite-elements. The Eulerian-Lagrangian formulation is used.

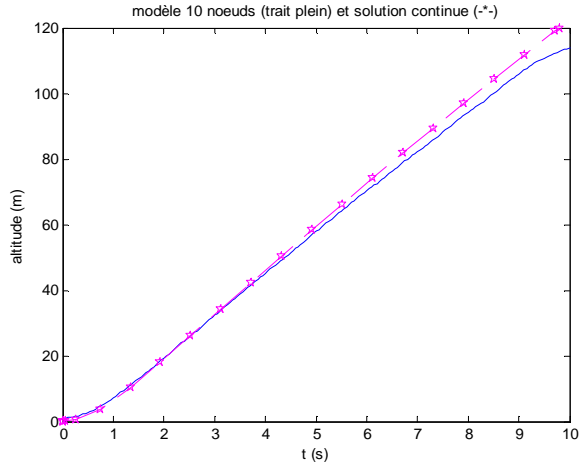

Figure 4: Exact (*) and finite-element (-) solutions of the drone altitude.

Figure 5 shows the velocity and the acceleration of the umbilical head-point. Note that the finite-element approximation gives numerical tension peaks during winch roll-out.

The time-dependant solutions are similar at the beginning of the take-off $t < 3$ s, later a difference appears and reaches 5% for the final altitude. We note significant discontinuities in velocity and acceleration in the discrete problem.

The numerical instabilities of the discrete problem follow each node unstacking. The difference between the models are significant after $t = 3$ s, near altitude $z = 30$ m. At that time, only two nodes fly over the winch. This observation suggests that the instability has a numeric cause. Note that it is favoured by higher umbilical stiffness and the absence of viscous dissipation provided by aerodynamic force. Considering a lateral wind, the instability is smoothed.

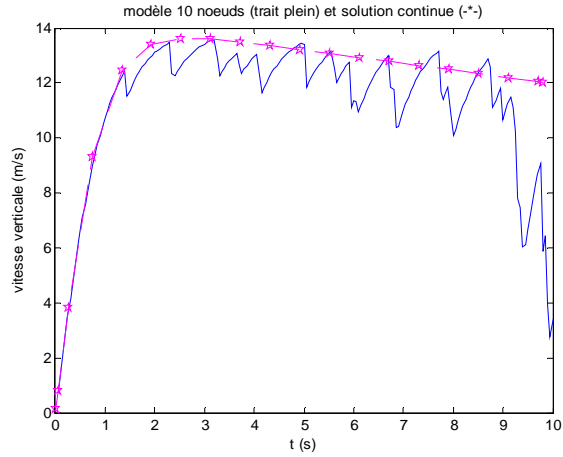

(a) Velocity.

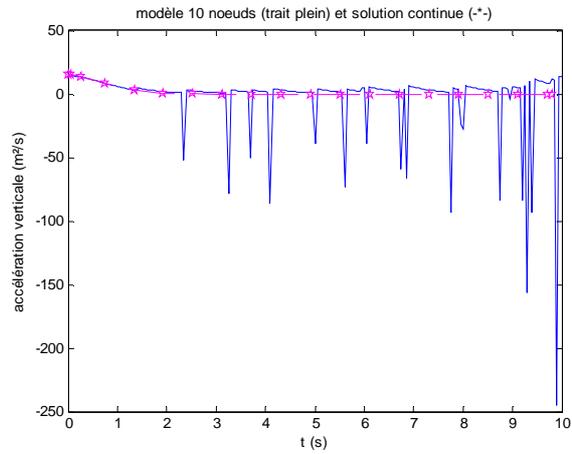

(b) Acceleration.

Figure 5: Drone take-off, exact (*) and finite-element solutions (-).

A node unstacking is combined with the brutal mass augmentation of the system. The bottom half part of the element located near the winch is related to that node weight. The mass steps of the system bottom imply that the acceleration of the umbilical head will be non-smooth. Our study initially assumed that the mass node at the winch was constant. Our result suggests introducing a time-dependant node mass assumption.

### 5.3 Transient loading

The section is based on the influence of wind. It illustrates a loading perturbation resulting in an equilibrium transition.

At the initial time, the system steady state corresponds to a normal flight phase. The umbilical is fully deployed. We study a modification of the wind velocity. It is assumed decreasing from a strong value to a small one. The other parameters are those of the previous section. The ship is fixed at the origin.

The wind is initially 10 m/s and goes to 5 m/s. The wind velocity follows a Heaviside function: $t = 0$, $V = 10$; $t > 0$, $V = 5$. The simulation duration is 30 s.

Figure 6 shows the transition with the positions of both drone and umbilical at following successive times: 0, 2.2, 4.3, 6.4, 8.5, 10.7, 12.8, 15, 17.2, 19.3 … 30 s. We determine a transient state during 20 s.

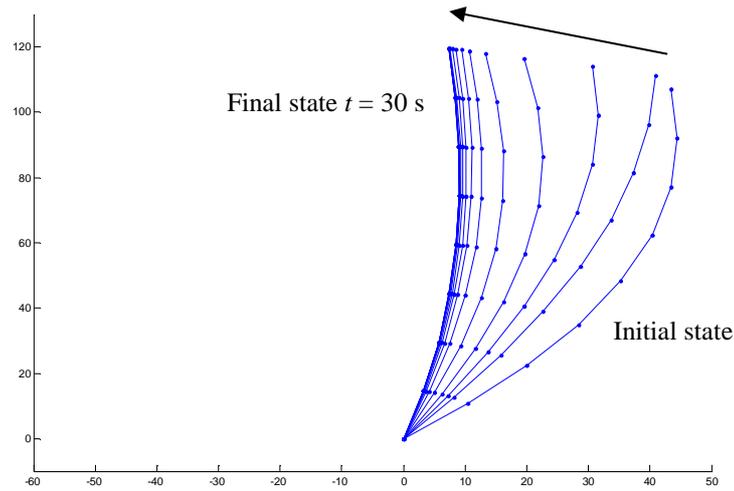

Figure 6: Time-evolution of the umbilical during a wind decrease.

With a reduced wind velocity, a reduced drag force acts on the umbilical. The lift-thrust force of the drone permits to stress significantly vertical the umbilical. The system moves upwind when the wind velocity decreases. The umbilical elongation is 25 cm with a 120 m total unstrained length.

Now, we study the inverse transition, when the wind velocity goes stronger from a small value. We consider the wind velocities 5 m/s and 10 m/s, the ship being immobile.

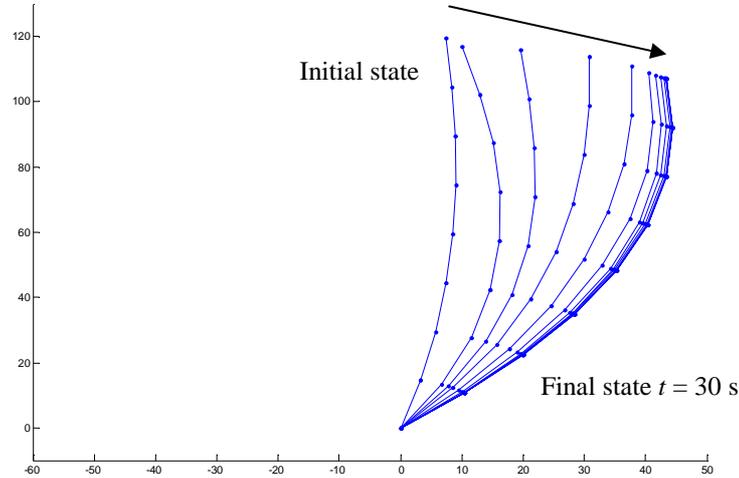

Figure 7:  Stronger wind velocity effect on the system.

In figure 7, to track the transient regime, the time interval is [0, 30] s. The discrete time values are the same that previously. This result demonstrates symmetric geometries with comparison of figures 6 and 7. Here, the transient duration is less. It equals 10 s indicative of higher viscous force. Therefore, the umbilical elongation is higher 28 cm than previously.

## 6   Conclusions

The numerical model proposed takes into account the umbilical elasticity. Generally, umbilical models use articulated rigid bodies. The model precision is expected sufficient to be consistent with the control-command performances of the system. The estimated weight respect and the authorized maximal altitude (149 m) are valuable objectives of future experimentations. A sequel of the research estimates the higher wind speeds and atmospheric conditions bearable by the system. Such system may have potential usages in agriculture, industry, natural disasters and coastal surveillance [14]. Standard oil resource rarefaction can potentially favour risky oil field exploitation resulting in further sea water pollution.

## Acknowledgments

This research was supported by the French Research Agency ANR, and ADEME, under the project RAPACE, grant n° ANR-05-ECOT-018-05. The author acknowledges the French frameworks, PRECODD ECOTECH and ECO-TS, on "Eco-Technologies and Eco-Services". The author thanks Mr. S. Nouchi and Dr. B. Variot for their valuable advices.